# On the influence of inversion on thermal properties of magnesium gallium spinel

*Linda Schwarz[1,2], Zbigniew Galazka[1], Thorsten M. Gesing[3], and Detlef Klimm*[*,1]*

Measurements of thermal diffusivity κ (using laser flash analysis) for temperatures from 100 °C to 1200 °C and of the inversion parameter *x* (using X-ray analysis) for the single crystal spinel $MgGa_2O_4$ have been performed. X-ray analysis showed a small change in *x* for samples annealed at 800 °C (*x* = 0.838) and 1200 °C (*x* = 0.834). Thermal diffusivity revealed a significant rise of κ for temperatures higher than 950 °C, deviating from the typical $1/T$ drop of κ. An additional systematic rise of κ with time is observed for temperatures between 950 °C and 1100 °C. A connection between cation disordering processes typical for spinels at elevated temperatures and the observed deviation of κ is proposed: Data indicate that cations changing their occupancy site at equilibration processes contribute significantly to κ and therefore represent a new type of heat transport.

In the so called *normal* spinel structure $A(B_2)O_4$, the cation A occupies 1/8 of the tetrahedral sites and the cation B 1/2 of the octahedral sites (marked by parentheses). $MgAl_2O_4$ is the most prominent example and crystallizes as normal spinel structure because $Mg^{2+}$ (71 pm) is much larger than $Al^{3+}$ (53 pm, Shannon's tetrahedral radii, [2]). In an *inverse* spinel $B(AB)O_4$, alternatively 1/8 of the tetrahedral sites are occupied by cation B whereas A and B occupy each 1/4 of the octahedral sites.

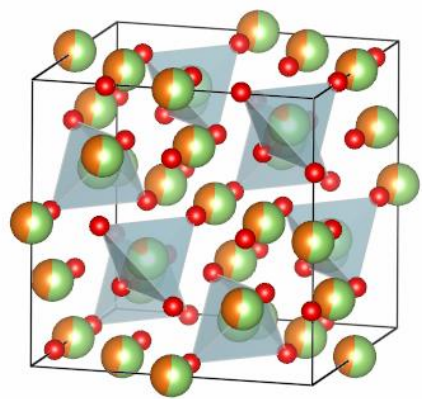

**Figure 1** Cubic $MgGa_2O_4$ spinel unit cell with Mg (orange), Ga (green) and O atoms (red) and blue painted tetrahedral sites. The ratio of partial colouring of Mg/Ga atoms stands for the site occupation factors of $MgGa_2O_4$, annealed at 1200 °C (4 h) and 800 °C (25 h) (see Table 1, sample MGO-800). Drawn with VESTA [3].

Cations A and B can rearrange – therefore, the normal and inverse spinel distributions represent extrema. The *degree of inversion x* characterizes the cation distribution in a more general formula $A_{1-x}B_x(A_xB_{2-x})O_4$.

## 1 Introduction

Spinels constitute a class of minerals of general type $AB_2X_4$, containing two cations A and B connected by anions X which is mostly oxygen. The spinel structure is cubic with space group $Fd\bar{3}m$. The unit cell contains 8 formula units, where 8 tetrahedral and 16 octahedral sites can be occupied either by cation A or B. The distribution of A and B ions depends on their charge as well as on the anion size. For 2-3 oxide spinels ($A^{2+}$, $B^{3+}$) the larger ions tend to occupy the tetrahedral site [1].

The parameter $x$ can show values between $x=0$ (normal spinel) and $x=1$ (inverse spinel).

In this study measurements with $MgGa_2O_4$ were performed. $Ga^{3+}$ (tetrahedral 61 pm) is still smaller than $Mg^{2+}$, but significantly larger compared to $Al^{3+}$, leading to an ordering which is called *mainly inverse* ($2/3<x<1$) [1]. It should be noted, however, that with increasing temperature $T$ the site occupancy, and thus the inversion degree $x$, becomes increasingly random [4]. A total randomness corresponds to $x=2/3$, which can be written as $A_{1/3}B_{2/3}(A_{2/3}B_{4/3})O_4$.

The increasing cation disorder of spinels at high temperatures goes along with increasing configurational entropy $S_c$ which can be expressed by

$$S_c = -R\bigl[x \ln x + (1-x)\ln(1-x) + x\ln\left(\frac{x}{2}\right) + (2-x)\ln\left(1-\frac{x}{2}\right)\bigr] \quad (1)$$

for $AB_2O_4$ spinels [1]. $S_c$ vanishes for $x = 0$, runs through a maximum $S_c \approx 15.88$ J/(mol K) at $x = 2/3$ and reaches $S_c = 2R\ln 2 \approx 11.53$ J/(mol K) for $x = 1$. Therefore the question occurs whether cation ordering influences the thermal diffusivity κ, defined by

$$\kappa = \frac{\lambda}{\rho\, c_p} \quad (2)$$

where $\lambda$ is thermal conductivity, $\rho$ is density (seen as constant) and $c_p$ is heat capacity at constant pressure. For insulators at high temperatures, $\lambda$ is influenced by photon scattering processes [5]. As $c_p$ changes only weakly for high temperatures (Dulong-Petit law), temperature dependence of κ goes with $\lambda$ and therefore with $T^{-1}$ [6].

If now also cation ordering contributes to the thermal conductivity, it should be dependent on temperature. So far, no measurements of thermal diffusivity have been performed for single crystal spinel phases at temperatures above 1000 °C. Observations on this subject are highly relevant for crystal growth experiments from a melt, where sufficient heat flow through the crystal is mandatory for a stable growth process.

The current study was motivated by peculiarities of differential scanning calorimetry (DSC) curves recently published (Fig. 1 in [7]). Galazka et al. [7] demonstrated that DSC curves showed an endothermal bend around 1000 °C, if the sample was hold prior to the measurement sufficiently long (several hours) at 800 °C. It was assumed that a change of the specific heat capacity $c_p$ resulting from cation rearrangement due to changes of the spinel inversion degree was the origin of the endothermal bend. In the current study the DSC results are amended by measurements of the thermal diffusivity in the same temperature region, and by single crystal X-ray diffraction studies on site occupation in the $MgGa_2O_4$ crystal structure.

## 2 Experimental

Recently, several melt growth techniques for $MgGa_2O_4$ single crystals were proposed [7]. For the current study samples (ca. 1.2 mm thickness and ca. 12 mm diameter for thermal diffusivity) were prepared from as-grown crystals obtained by the Kyropoulos method from an iridium crucible and in argon atmosphere.

Thermal diffusivity was measured by laser flash technique (LFA 427, Netzsch). The laser generates heat pulses on the bottom of the sample disc, dissipating to the top surface of the sample, where the time-dependent signal is measured and evaluated: the half rise time is used to fit the signal, considering passing radiation [8].

Measurements were performed at ambient pressure in nitrogen atmosphere. The latter was necessary to prevent oxidation of a graphite layer which was sprayed onto the surfaces prior to each measurement to optimize absorption of the laser flash and to ensure opacity. Other trials were performed in pure oxygen, to enforce stability of $Ga^{3+}$ with respect to reduction to metallic Ga. For these measurements instead of carbon a 50 nm coat of platinum was evaporated onto the surfaces. Surprisingly the Pt layer vanished after exposure to high temperatures. It is not clear whether the metal diffused into the $MgGa_2O_4$ crystal volume, possibly after oxidation by the $O_2$ atmosphere, or evaporated. It should be noted, however, that the fugacity of platinum (as $PtO_2$) is only at the level of $1.5\times10^{-6}$ bar at 1200 °C [9] – such a low fugacity makes evaporation not very feasible. The problems connected with the creation of stable absorption layers restricted the accessible temperature range to ca. 1150 °C.

During the measurements, the $MgGa_2O_4$ sample was placed on a ceramic $Al_2O_3$ aperture of 6 mm diameter which was therefore also used as the sample's diameter.

The temperature range of the measurements was between 100 °C and 1150 °C. For each measurement point the sample was heated to the desired temperature value (5 degrees/min for $T\leq400$ °C; 10 degrees/min for $T\geq500$ °C), tolerating a temperature accuracy of ±3 degrees. A series of laser shots, with parameters adjusted to each measurement, was then performed. The laser voltage was adapted to the sample and best results, with respect to sample stability and signal quality, were obtained using $U = 460$ V and a pulse width $t = 0.4$ ms.

For X-ray diffraction experiments two pieces (each ca. 50 mg) were annealed in a Netzsch STA 409C DTA apparatus at 1200 °C (4 hours) and 1200 °C/800 °C (4 hours/25 hours), respectively. After these treatments the samples were cooled as fast as possible (maximum rate ca. 60 degrees/min) to room-temperature. Unfortunately it proved to be impossible to maintain such high cooling rate below 600 – 700 °C due to the large thermal inertia of a DTA furnace.

For single-crystal X-ray diffraction data collection, suitable crystals were selected and mounted on the top of a small glass fibre fixed within a metal pin. Single crystal X-ray diffraction was performed on a Bruker AXS D8 Venture equipped with a KAPPA four-circle goniometer. The PHOTON 100 detector based on CMOS technology provided an active area of 100 cm$^2$. The experiments were carried out using Mo K$\alpha$ ($\lambda$ = 71.073 pm) radiation. A curved graphite crystal TRIUMPH served as monochromator. A semi-empirical approach was applied for absorption correction for equivalent reflections, leading to a $R_{int} \sim 0.06$ for both crystals. During the refinement using Shelx-97 [10] with anisotropic displacement parameters for all atoms a total of 10 parameters were varied resulting in final residuals of $R_1 = 0.0097$ for 133 $F_0 > 4\sigma F_0$, and 0.0107 for all 141 data for the crystal treated at 800 °C. For the other crystal treated at 1200 °C, $R_1 = 0.0114$ for 139 $F_0 > 4\sigma F_0$ and 0.0115 for all 141 data. The refined lattice parameter, atomic coordinates, anisotropic and equivalent isotropic displacement obtained from the room-temperature data are given in Table 1. The mixed occupancy of the cations on the *8a* and *16d* position of space group $Fd\bar{3}m$ was refined assuming the constraint of full occupancy of both sides. Additionally the displacement parameters for each individual side were constrained during the refinement.

Table 1. Crystallographic data of Mg$_{1.07(2)}$Ga$_{1.93(2)}$O$_{4-\delta}$ at 800 °C and Mg$_{1.08(2)}$Ga$_{1.92(2)}$O$_{4-\delta}$ at 1200 °C (italics). The occupancy of Ga on Wyckoff position 8a is the inversion degree *x*.

| 800 °C, $Fd\bar{3}m$: $a$ = 828.07(3) pm. $R_{int}$ = 0.0617, $R_1$ = 0.0097, $R_{1(all)}$ = 0.0107, $wR_2$ = 0.0235, GooF = 1.424[a] | | | | | | |
|---|---|---|---|---|---|---|
| *1200 °C, $Fd\bar{3}m$: $a$ = 828.25(5) pm. $R_{int}$ = 0.0610, $R_1$ = 0.0114, $R_{1(all)}$ = 0.0115, $wR_2$ = 0.0267, GooF = 1.509* [a] | | | | | | |
| Atom | Wyckoff | Site sym. | Occupancy | x | y | z |
| Ga/Mg1 | 8a | $\bar{4}3m$ | 0.838(8)/0.162(8) | ⅛ | ⅛ | ⅛ |
|  |  |  | *0.834(8)/0.166(8)* |  |  |  |
| Ga/Mg2 | 16d | $.\bar{3}m$ | 0.548(6)/0.452(6) | ½ | ½ | ½ |
|  |  |  | *0.541(6)/0.459(6)* |  |  |  |
| O1 | 32e | $.3m$ | 1 | 0.25672(5) | x | x |
|  |  |  | *1* | *0.25674(6)* | *x* | *x* |
| Anisotropic displacement parameters ×10$^4$/pm$^2$ | | | | | | |
| Atom | $U_{11}$ | $U_{22}$ | $U_{33}$ | $U_{12}$ | $U_{13}$ | $U_{23}$ | $U_{eq}$ |
| Ga/Mg1 | 0.0038(1) | $U_{11}$ | $U_{11}$ | 0 | 0 | 0 | 0.0038(1) |
|  | *0.0039(1)* |  |  |  |  |  | *0.0039(1)* |
| Ga/Mg2 | 0.0040(1) | $U_{11}$ | $U_{11}$ | -0.0005(1) | $U_{12}$ | $U_{12}$ | 0.0040(1) |
|  | *0.0040(1)* |  |  | *-0.0005(1)* |  |  | *0.0040(1)* |
| O1 | 0.0064(2) | $U_{11}$ | $U_{11}$ | -0.0002(1) | $U_{12}$ | $U_{12}$ | 0.0064(2) |
|  | *0.0064(2)* |  |  | *-0.0001(1)* |  |  | *0.0064(2)* |

[a] 13098 and 12823 data were collected in the range of -16 ≤ h, k, l ≤ 16 for the 800 °C and 1200 °C sample, respectively, resulting in 141 unique reflections each.

## 3 Results and Discussion

Measurements of thermal diffusivity were first made by simply heating the sample from 100 °C to 1150 °C. A laser shot series of 3 shots and waiting time of 1 minute between the shots every 100 K, or from 900 °C every 50 K with 5 laser shots, was performed (see Fig-ure 2). For temperatures until 800 °C, the typical dependence $\kappa \propto T^{-1}$ characteristic for heat conductivity of insulators at elevated temperatures can be observed [5]. Reaching temperatures of 1000 °C and higher, a significant rise of κ is seen. It is interesting to note that the experimental data by Wilkerson et al. (Fig. 2 in [11]), obtained with ceramic $MgAl_2O_4$-$MgGa_2O_4$ solid solutions, indicate for some samples a similar positive κ (*T*) deviation near 1200 °C. However, this effect was less significant with the polycrystalline samples that were used by these authors, compared to the data presented here in Figure 2. They were not discussed by Wilkerson et al. [11].

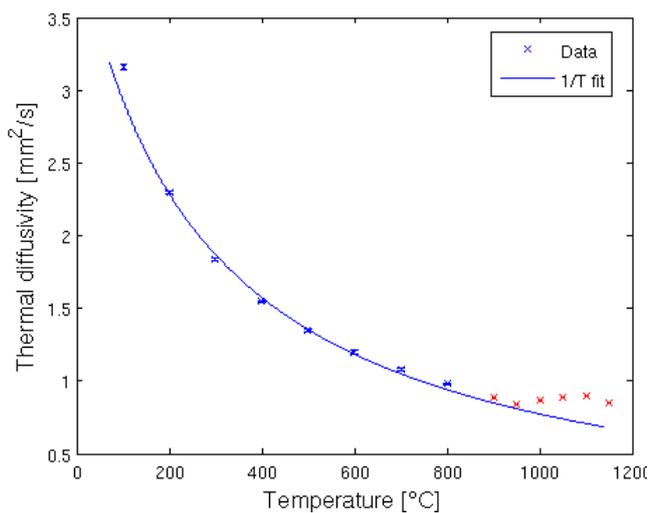

**Figure 2** The thermal diffusivity κ vs temperature *T* of a $MgGa_2O_4$ single crystal. Each value represents the average of 3-5 measurements, the uncertainty consists of their discrepancy. The sample was heated from 100 °C to 1150 °C within 330 minutes. A weighted 1/*T* fit was made with values coloured in blue. Red coloured values visibly deviate from that fit, probably indicating new heat transport processes in the spinel.

The positive deviation of the κ (*T*) curve appears in the same *T* range as the endothermal bend in the DSC curves that was mentioned in ref. [7]. In a similar *T* range, close to 1000 °C, significant changes of the inversion degree were observed for $MgAl_2O_4$ [12] and $CuAl_2O_4$ [13], respectively. It seems to be a realistic hypothesis that mutual exchange of ions between tetrahedrally and octahedrally coordinated sites is responsible for the endothermal bend of DSC curves [7] as well as increased thermal diffusivity. This means that not only phonons and electrons, but also moving ions can contribute significantly to the transport of heat in crystals.

Such diffusive steps require time. To confirm this, in additional measurements, the spinel was first hold at 800 °C for 1 h, giving time for supposed equilibration processes. Thereafter 6 laser shots with waiting time of ca. 4 minutes between each of them were performed for temperatures between 900 °C and 1150 °C in 50 K steps (Figure 3). Thermal diffusivity now reached a maximum approximately 300% with respect to the maximum value reached by simply heating the sample. The systematic time-dependency may indicate that cation ordering still takes place.

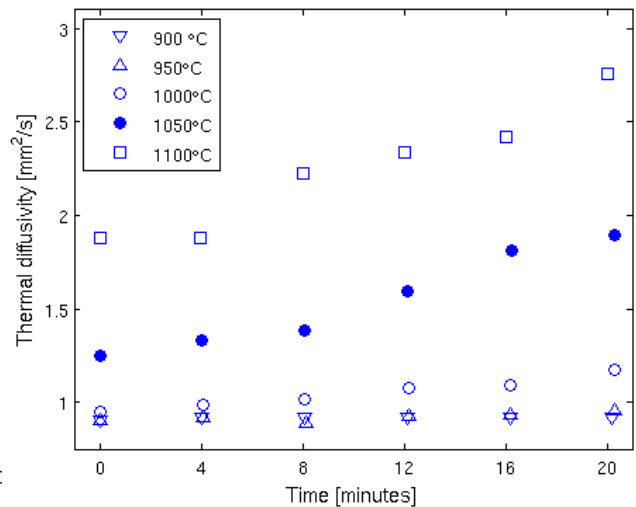

**Figure 3** The thermal diffusivity κ vs time of single crystal of $MgGa_2O_4$. The spinel was hold at 800 °C for one hour and then was successively heated. `0 minutes' corresponds to the first measurement at the indicated temperature. A time dependant rise of κ with the temperature can be observed.

In the introduction it was discussed that the degree of inversion approaches for all spinels the value *x* = 2/3 at sufficiently high temperature. This value means for $MgGa_2O_4$ random ordering of $Mg^{2+}$ and $Ga^{3+}$ on the tetrahedral or octahedral sites, respectively. By X-ray diffraction studies with a single crystalline sample which was equilibrated at 1200 °C or 800 °C, respectively, it was investigated if different site distributions prevail also after cooling to room temperature. However, the results in Table 1 show that the differences between both samples are only marginal. The structure drawing in Figure 1 represents the data for the MGO-800 sample, but differences for MGO-1200 are so small that they would not be recognized visually.

Duan et al. [14] report for $MgGa_2O_4$ nanocrystals in their Table 4 a fraction of 0.21 $Ga^{3+}$ on the tetrahedral site, but in contrast they give an inversion degree 0.45 for the same samples in Table 5. One can suspect that this discrepancy is related to their unclear definition of $x$.

The inversion degree $x$ measured in the current study is the fraction of tetrahedral sites occupied by $Ga^{3+}$ ions, which is $x=0.834(8)$ (annealing at 1200 °C) or $x=0.838(8)$ (annealing at 800 °C) for both samples of this study. This difference is small, which might result from too slow cooling after the annealing; but it is in the expected direction, closer to random order $x = 2/3$ for MGO-1200. The only marginal different inversion degree results with equation (1) in a very minor entropy difference of $\Delta S = 0.043$ J/(mol K) between both samples.

However, under the given experimental conditions it seems more realistic to assume that at high temperature (this means above ≈1200 °C) ion ordering is indeed random; which means $S_c \approx 15.88$ J/(mol K). If this assumption holds, one calculates $\Delta S = 0.8865$ J/(mol K). At the observed transformation temperature, $T_t \approx 1400$ K, this entropy change corresponds to a Gibbs energy reduction of $MgGa_2O_4$ of $T_t \Delta S \approx 1.24$ kJ/mol, which means that the spinel phase experiences entropic stabilization.

Unfortunately, the recently measured $c_p$ changes [7] up to 0.06 J/g K = 13.7 J/(mol K) cannot be correlated straightforward with the entropy changes mentioned above. Enthalpy as well as entropy contribute to $\Delta c_p$, and even for proteins both contributions can be significant [15]. It must be expected that with ionic bonding, like for $MgGa_2O_4$, enthalpic contributions are even larger.

## 4. Conclusion

Measurements of thermal diffusivity κ of single crystal $MgGa_2O_4$ showed a significant rise in κ for temperatures from 1000 °C and above. When these temperatures are kept constant after holding the sample at 800 °C for 1 hour, a time-dependant rise of κ is observed. X-ray analysis of the inversion parameter $x$ indicates a small change in $x$ directing to random distribution of cations. These observations, as well as earlier studies connecting $x$ with temperature $T$ and time for equilibration processes [4],[13], lead to the assumption of a new form of heat transport: Cations changing their site by disordering processes might contribute to thermal conductivity related with κ.

To the best of our knowledge, no correlation between cation disorder processes and heat transport has been reported so far. Shukla et al. [16] simulated a correlation between thermal conductivity and $x$ for the spinel $MgAl_2O_4$ and found a very weak dependency, caused by phonon scattering from anti-site effects. As these results do not fit to the presented observations, further theoretically investigations in heat transport generated by cation disordering are encouraged. The remaining questions are how thermal diffusivity changes with temperatures $T>1200$ °C and if the observed characteristics of $MgGa_2O_4$ are specific for this single crystal or if they can be generalized for a specific group or even all spinels.

**Acknowledgements**.

The authors thank Martina Rabe and Mike Pietsch for sample preparation. We express our gratitude to Reinhard Uecker for reading the manuscript. TMG gratefully acknowledges the Deutsche Forschungsgmeinschaft (DFG) for the financial support in the Heisenberg program (GE1981/3-1 and GE1981/3-2) and Malik Šehović (CKfS, Uni Bremen) for the single crystal data collection.

* Corresponding author: e-mail detlef.klimm@ikz-berlin.de, Phone: +49-30-6392-3018, Fax: +49-30-6392-3003

[1] Leibniz-Institut für Kristallzüchtung, Max-Born-Str. 2, 12489 Berlin, Germany

[2] Humboldt-Universität zu Berlin, Institut für Physik, Newtonstr. 15, 12489 Berlin, Germany

[3] Universität Bremen / FB 2; Chemische Kristallographie fester Stoffe; Institute für Anorganische Chemie und Kristallographie und MAPEX Center for Materials and Processes; Leobener Straße / NW2, D-28359 Bremen, Germany